\documentclass[twocolumn]{aastex631}

\usepackage{textcomp}
\usepackage{amsmath}
\usepackage{float}

\received{2021 April 20}
\revised{2021 September 21}
\accepted{2021 November 9}

\begin{document}

\title{Lifting of Tribocharged Grains by Martian Winds}

\correspondingauthor{Maximilian Kruss}
\email{maximilian.kruss@uni-due.de}

\author{Maximilian Kruss}
\affil{University of Duisburg-Essen, Faculty of Physics, Lotharstr. 1-21, 47057 Duisburg, Germany}

\author{Tim Salzmann}
\affil{University of Duisburg-Essen, Faculty of Physics, Lotharstr. 1-21, 47057 Duisburg, Germany}

\author{Eric Parteli}
\affil{University of Duisburg-Essen, Faculty of Physics, Lotharstr. 1-21, 47057 Duisburg, Germany}

\author{Felix Jungmann}
\affil{University of Duisburg-Essen, Faculty of Physics, Lotharstr. 1-21, 47057 Duisburg, Germany}

\author{Jens Teiser}
\affil{University of Duisburg-Essen, Faculty of Physics, Lotharstr. 1-21, 47057 Duisburg, Germany}

\author{Laurent Schönau}
\affil{University of Duisburg-Essen, Faculty of Physics, Lotharstr. 1-21, 47057 Duisburg, Germany}

\author{Gerhard Wurm}
\affil{University of Duisburg-Essen, Faculty of Physics, Lotharstr. 1-21, 47057 Duisburg, Germany}

\begin{abstract}
It is a long-standing open question whether electrification of wind-blown sand due to tribocharging --- the generation of electric charges on the surface of sand grains by particle-particle collisions --- could affect rates of sand transport occurrence on Mars substantially. While previous wind tunnel experiments and numerical simulations addressed how particle trajectories may be affected by external electric fields, the effect of sand electrification remains uncertain. Here we show, by means of wind tunnel simulations under air pressure of 20\,mbar, that the presence of electric charges on the particle surface can reduce the minimal threshold wind shear velocity for the initiation of sand transport, $u_{{\ast}{\mathrm{ft}}}$, significantly. In our experiments, we considered different samples, a model system of glass beads as well as a Martian soil analog, and different scenarios of triboelectrification. Furthermore, we present a model to explain the values of $u_{{\ast}{\mathrm{ft}}}$ obtained in the wind tunnel that is based on inhomogeneously distributed surface charges. Our results imply that particle transport that subsides, once the wind shear velocity has fallen below the threshold for sustained transport, can more easily be restarted on Mars than previously thought.
\end{abstract}


\section{Introduction}

Mars has an atmosphere that is almost two orders of magnitude lower in surface pressure than Earth's, and nevertheless Martian dunes are migrating daily and producing Earth-like sand flux rates, while rovers and satellites have been detecting dust devils and global dust storms on the red planet \citep{Fisher2005, Bridges2012, Bourke_et_al_2019,Liuzzi2020, Heyer2020}. Indeed, while a breeze is sufficient to lift sand grains from the ground at 1\,bar on Earth, it requires a storm to do the same at a few mbar $\rm CO_2$ on Mars \citep{Bagnold1941, Greeley1985, Merrison2007, Rasmussen2015, Burr2020}. This makes conditions for particle lifting more critical on the Martian surface, as it might well depend on minute details if a region is active or inactive with respect to particle transport and if dust entrainment into the atmosphere is possible or not.

The physics of soil particle entrainment by wind forces is still today a matter of research \citep{Paehtz_et_al_2020,Fu_2020}. There have been a number of ideas to explain how dust and sand can be moved more easily on Mars \citep{Neakrase2016}. Previous experiments showed that dust aggregates are more easily mobilized than sand \citep{Merrison2007} and that sand can be lifted from the soil in sporadic events of strong winds on the surface of Mars \citep{Swann2020}. Furthermore, the low Martian gravity also seems to be beneficial to particle lifting, since granular material settled under the gravity of Mars produces granular packings associated with lower solid fractions than their Earth counterparts \citep{White1987, Musiolik2018, Kruss2020}. 

However, one further ingredient in the physics of soil grain entrainment on Mars is the rarified Martian atmosphere. Recently, \citet{Demirci2020a} showed that the transition from hydrodynamic flow to molecular flow around an individual grain increases the necessary shear stress for smaller particles further. For a less penetrable ground of small grains, lifting might be easier owing to a pressure difference between subsoil and above-ground pressure as vortices travel the surface \citep{Greeley2003, Balme2006, Neakrase2016, Koester2017a, Bila2020}. Nevertheless, a low pressure might not always be detrimental to particle lifting. As the flow becomes molecular through the pores of the soil, thermal creep gas flow sets in, following temperature gradients from cold to warm \citep{Koester2017}. This can lead to a subsoil overpressure on the insolated Martian surface \citep{deBeule2014, Kuepper2016, Schmidt2017}.
Summarizing, these ideas on lifting turn the screws on the conventional forces acting on grains: gas drag and pressure, gravity, and adhesion. However, there is at least one more force that needs to be considered in this context, which is electrostatics \citep{Rasmussen2009, Esposito2016}.

It is known that sand storms can produce large electric fields of the order of 100\, kV/m close to the ground \citep{Schmidt1998, Zheng2013, Zhang2020}. In addition, dust devils show electric fields \citep{Franzese2018}. Volcano eruptions regularly come with lightning as a visible sign of charging and charge separation \citep{Aplin2014, Harper2016}. The respective fields are strong enough to pick up charged grains. The effectiveness of such electric fields in lifting sand grains has been tested, for instance, through modeling and experiments \citep{Kok_and_Renno_2006,Paehtz_et_al_2010}. Specifically, this previous work considered that the grains are well conducting and subject to an external field, which induces charges on the grains, thus producing an upward-pointing electrostatic force \citep{Kok_and_Renno_2006}. 

While this previous work relied on an externally applied electric field, sand-driven electric fields are certainly self-generated by triboelectric effects as the sand grains are hopping and colliding with each other within the transport layer \citep{Wurm2019, Harper2021}. However, it is by no means clear whether triboelectric charging lowers the minimal threshold wind shear velocity for transport initiation, $u_{{\ast}{\mathrm{ft}}}$, since charges may promote aggregation, thus making particle lifting more difficult \citep{Harper2017, Steinpilz2020a, Steinpilz2020b, Teiser2021}.

Here, we thus investigate the influence of tribocharging on $u_{{\ast}{\mathrm{ft}}}$ by means of wind tunnel experiments operating at Martian-like atmospheric pressure conditions (20\,mbar). In contrast to previous work, we do not apply any external electric field to the granular bed. Instead, we use particle collisions to charge a granular bed, whereupon we analyze the effect of this charge on $u_{{\ast}{\mathrm{ft}}}$. Moreover, we assume that the ground is not able to conduct charges. As fields exist, a conductive discharge does not occur on relevant timescales of seconds to minutes at least. Our assumption is thus plausible, as regions of sand activity and especially the surface of Mars are rather arid, with low humidity making sand grains extremely bad conductors.  Moreover, the charges of the particles within the soil provide the source of the electric fields acting on a particle that is at rest on the surface.

\section{Experiment}

In this section we describe the experimental setup developed to simulate the entrainment of particles under Martian-like pressure conditions, the granular material employed in our experiments, and the methods for charging and discharging of this granular material.

\subsection{Experimental setup and the wind tunnel}

\begin{figure}
 \centering 
\includegraphics[width=1.0\columnwidth]{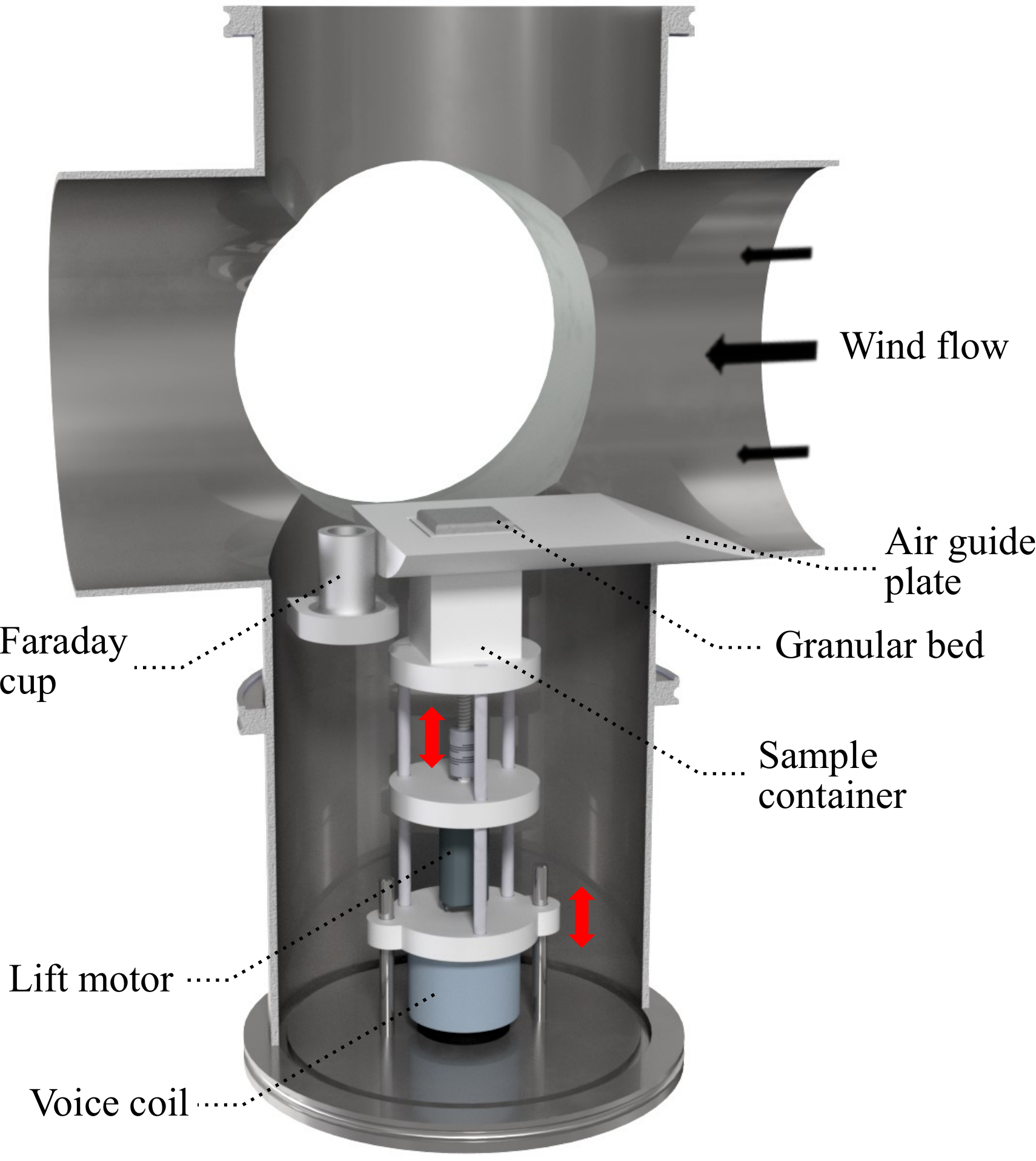}
\caption{Setup of the observation chamber. A charging unit is integrated into a low-pressure wind tunnel, while in the experiments simulating uncharged grains, a discharge unit is mounted on the upper flange.}
\label{fig:aufbau}
\end{figure}

The main component of the experimental setup is a wind tunnel with an inner tube diameter of 6.3\,cm. The maximum Reynolds number inside the wind tunnel is on the order of $\rm{Re}=500$, but the exact wind profile is not known. Fig.~\ref{fig:aufbau} shows the observation cell, which is located in a horizontal section of the wind tunnel and hosts the granular bed. This cell is equipped with viewports at its front and back sides, which enable visualization of the granular dynamics with the camera through one side and illumination through the other side. The entire wind tunnel acts as a vacuum chamber to simulate a thin atmosphere such as the one of Mars.

The experiment is observed with a high-speed camera, which is combined with an LED panel as bright field illumination (in backlight configuration) for a visualization of the single grain trajectories. In our experiments, the camera operates at 1000\,fps with an exposure time of 1/5000\,s, a focal depth of 1\,mm, and a spatial resolution of 25\,\textmu m.

The wind tunnel operates at 293\,K and an air pressure of 20\,mbar. This air pressure is the lowest value suitable for our experiments. If we further decrease the air pressure to simulate the mean atmospheric pressure on the Martian surface (6\,mbar), then the roots pump of our wind tunnel is not able to produce strong enough surface winds for causing erosion of the granular bed under Earth's gravity. Therefore, all results in the present work have been obtained using an atmospheric pressure of 20\,mbar, or equivalently with an air density of $0.023\,$kg/m$^3$.

\subsection{Granular material}

The granular bed measures 3\,cm~x~3\,cm with a thickness of about 1\,cm in all experiments. As a granular sample we choose two different materials. Monodisperse soda-lime glass beads of a mean diameter of approximately 434\,\textmu m with a standard deviation of $\pm\,$17\,\textmu m serve as model particles that allow a theoretical model for charge-modified sand entrainment to be created. For the exact size distribution we refer to \citet{Steinpilz2020a}. After vibrating the sample as described in Section~\ref{sec:charging}, a packing fraction $\phi$ of the bed of around 0.60 was measured.

\begin{figure}
 \centering 
\includegraphics[width=1.0\columnwidth]{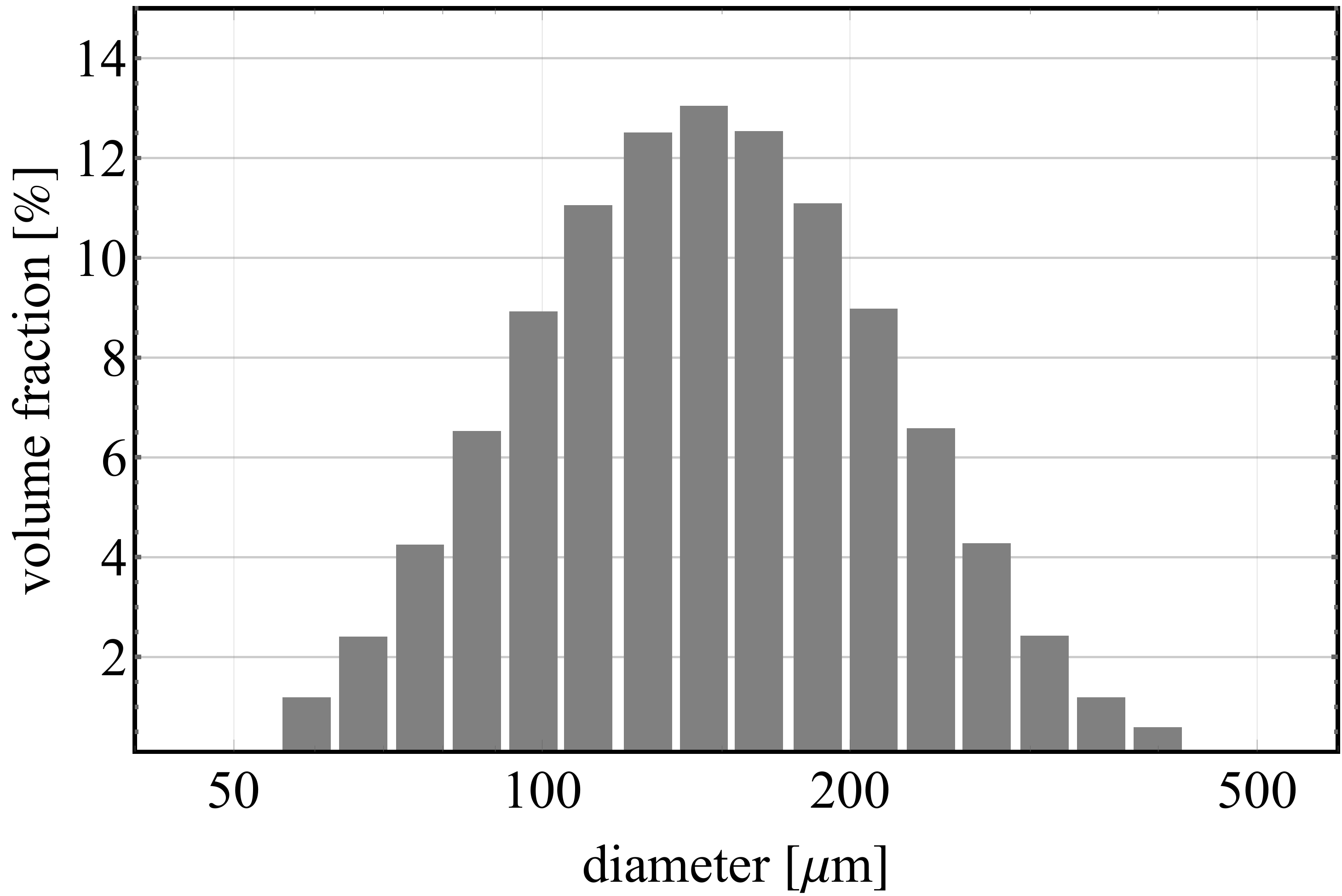}
\caption{Grain size distribution of the used sample of MGS.}
\label{fig:sizedistribution}
\end{figure}

In addition, Mars Global Simulant (MGS), a mineralogical analog to Martian regolith, is used to evaluate the effect in a more realistic scenario. This material was characterized in detail by \citet{Cannon2019}. The size distribution of the sand grains composing the sample is shown in Fig.~\ref{fig:sizedistribution}.

\subsection{\label{sec:charging}Charging}

To generate electric charges on the sample particles (tribocharging), the granular material is placed into a 3D-printed container of PETG plastic, which is subjected to controlled agitation produced by a voice coil. While the glass samples are agitated in an uncoated plastic container, the sample container for the sand measurements is covered with MGS grains to ensure the same material contacts. To avoid spilling of the sample, the granular bed can be lowered in the container for the duration of the shaking procedure and then moved up again using the lift motor shown in Fig.~\ref{fig:aufbau}. We perform different types of agitation:

\begin{itemize}
\item {\em{Charged beads \#1}}: The glass beads are {\em{circulated}} for 2\,hr in the container by suitably setting the frequency of the voice coil to 19\,Hz with a stroke of around 1\,mm (peak to peak). In this case, the particles move down at the container walls and then up again in the bulk material.

\item {\em{Charged beads} \#2}: The glass beads are {\em{shaken}} for 2\,hr by applying controlled vibrations to the container with an amplitude of around 1\,mm and a frequency of 15\,Hz. At this frequency, the particles in the top layers bounce on the surface without being mixed into the bulk material.

\item {\em{Charged sand}: The MGS sample is {\em{shaken}} for 1~minute, which was found to be sufficient to saturate the charges, with an amplitude of around 1\,mm and a frequency of 17\,Hz.}

\end{itemize}

Furthermore, to quantify the role of tribocharging on particle entrainment, we need to compare the results obtained with the charged beds with the outcomes from experiments performed without charging. In that case, the granular bed is not shaken prior to the experiment. To reduce the charges as much as possible, the glass beads and the sand sample are subjected to a plasma surface flow (overall neutral) using an air ionizer that is directed onto the granular bed from above. Thereafter, the system is closed and evacuated to 20\,mbar.

\subsection{Experimental procedure for estimating the minimal threshold wind shear velocity $u_{{\ast}{\mathrm{ft}}}$}

The goal is to identify the minimal flow speed in the wind tunnel that is just sufficient to lift off single particles from the surface. We use \textit{Trackmate} as an automated particle tracker \citep{Tivenez2017}, which allows us to measure the position and the velocity of moving glass beads in the system at every 1\,ms, based on the frame rate of 1000\,fps in the experiments. Due to their irregular shapes and wide size range, the sand grains are tracked manually.

The wind velocity at different heights above the ground is estimated from the trajectories of the hopping particles. Specifically, the following equation gives the horizontal position $x(t,h)$ of a grain at time $t$ and height $h$ above the surface \citep{Wurm2001},

\begin{equation}
x(t,h) = \left[{u(h)-v_{0,x}}\right]\cdot \tau \cdot {\mbox{exp}}{\left({-\frac{t}{\tau}}\right)}+u(h) \cdot t + x_0 \label{eq:x}
\end{equation}

where $u(h)$ denotes the wind velocity at height $h$, $v_{0,x}$ and $x_0$ are the particle lift-off velocity and position, respectively, and $\tau$ is the coupling time, i.e., the characteristic time needed by the particle to adapt to the wind velocity.

It was shown by \cite{Merrison2008} that, under the air pressure conditions associated with our experiments, the air flow near the granular bed is laminar, so that the wind shear velocity can be estimated from the height profile of the wind velocity with the equation

\begin{equation}
u_{\ast} = \sqrt{\frac{\eta}{\rho_{\mathrm{f}}}\frac{{\mathrm{d}}u(h)}{{\mathrm{d}}h}}, \label{eq:ustar}
\end{equation}

where $\eta=18\cdot 10^{-6}\,\rm{Pa}\cdot\rm{s}$ is the dynamic viscosity of the air in the wind tunnel, $h$ is the height above the surface, and $\rho_{\mathrm{f}}$ is the air density.

\subsection{Estimating the particle charge in the experiments}

In order to measure the charge of the particles, a Faraday cup is suitably positioned at the downwind end of the observation cell to collect the particles hopping along the bed. This Faraday cup is connected to an electrometer, which allows us to measure the charge of the grains. To this end, experiments are performed at the minimum threshold shear velocity $u_{{\ast}{\mathrm{ft}}}$, so that only one particle at a time enters the cup, thus enabling measurements of the charges of single particles. For details on the charge measurement we refer to \citet{Genc2019} and \citet{Jungmann2021}. Each experiment is performed for about 10\,minutes and no particles with a charge magnitude smaller than $0.2 \cdot 10^5\,$e can be detected. We note that the charge measurement can only be carried out for the model system of glass beads, as it requires well-defined, monodisperse particles. 
Since it is not possible to determine the number of sand grains entering the cup at a time, we cannot give charges of individual sand grains.

\section{Experimental results}

\subsection{Charge measurements}

\begin{figure}
 \centering 
\includegraphics[width=1.0\columnwidth]{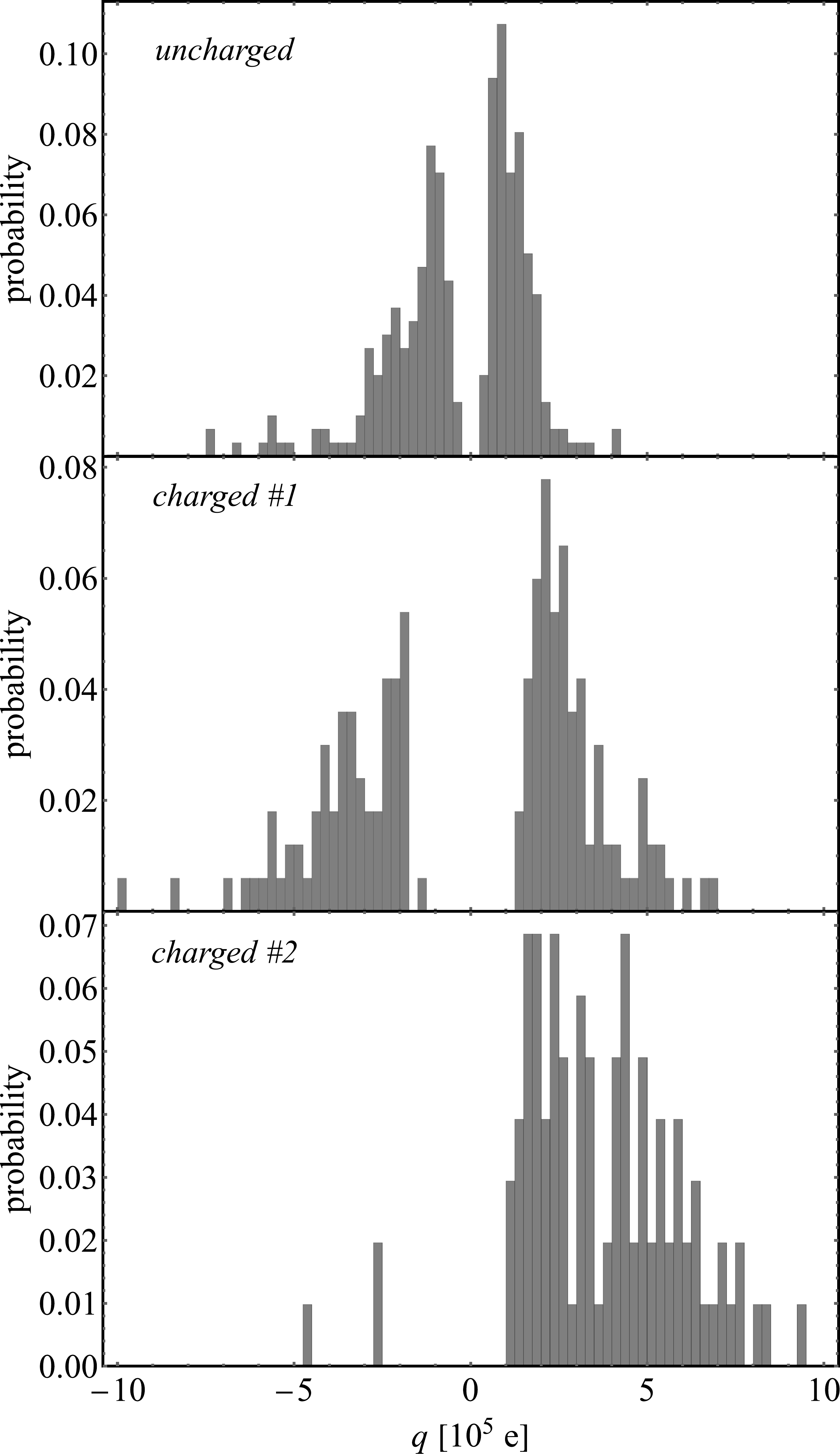}
\caption{Measured charges of glass beads for \textit{uncharged} and both other cases.}
\label{fig:daten_charge}
\end{figure}

The charge distribution of the glass beads in all three scenarios (\textit{uncharged, charged \#1, charged \#2}) is displayed in Fig.~\ref{fig:daten_charge}. The charge $q$ of the particles is given in units of elementary charges of ${\mbox{e}}\approx 1.602 \cdot 10^{-19}\,$C.

It has to be noted that the \textit{uncharged} case still shows significant charges. We attribute this to the motion of lifted beads that are not entrained by the wind but fall back to the surface. Those particles hop over the bed surface and charge before reaching the Faraday cup. In the \textit{charged} cases, the shaking procedure results in broader charge distributions with almost up to $10^6$\,e peak charge. Depending on the type of agitation, one charge distribution is almost symmetrical, while the other contains predominantly positive charges, which can be attributed to the dielectric difference between the glass particles and the plastic container. The gaps in the distributions around $q=0$ are due to the limited resolution of the electrometer.

Although we cannot measure the charge distributions of the sand particles, these are assumed to be symmetrical, as charges are generated by the same material tribocharging in the coated container. If the results of previous experiments using larger grains are scaled to the sizes shown in Fig.~\ref{fig:sizedistribution}, charges of $10^5$\,e to $10^6$\,e can be expected \citep{Wurm2019}.

\subsection{Minimal threshold wind shear velocity}

\begin{figure}
\includegraphics[width=\columnwidth]{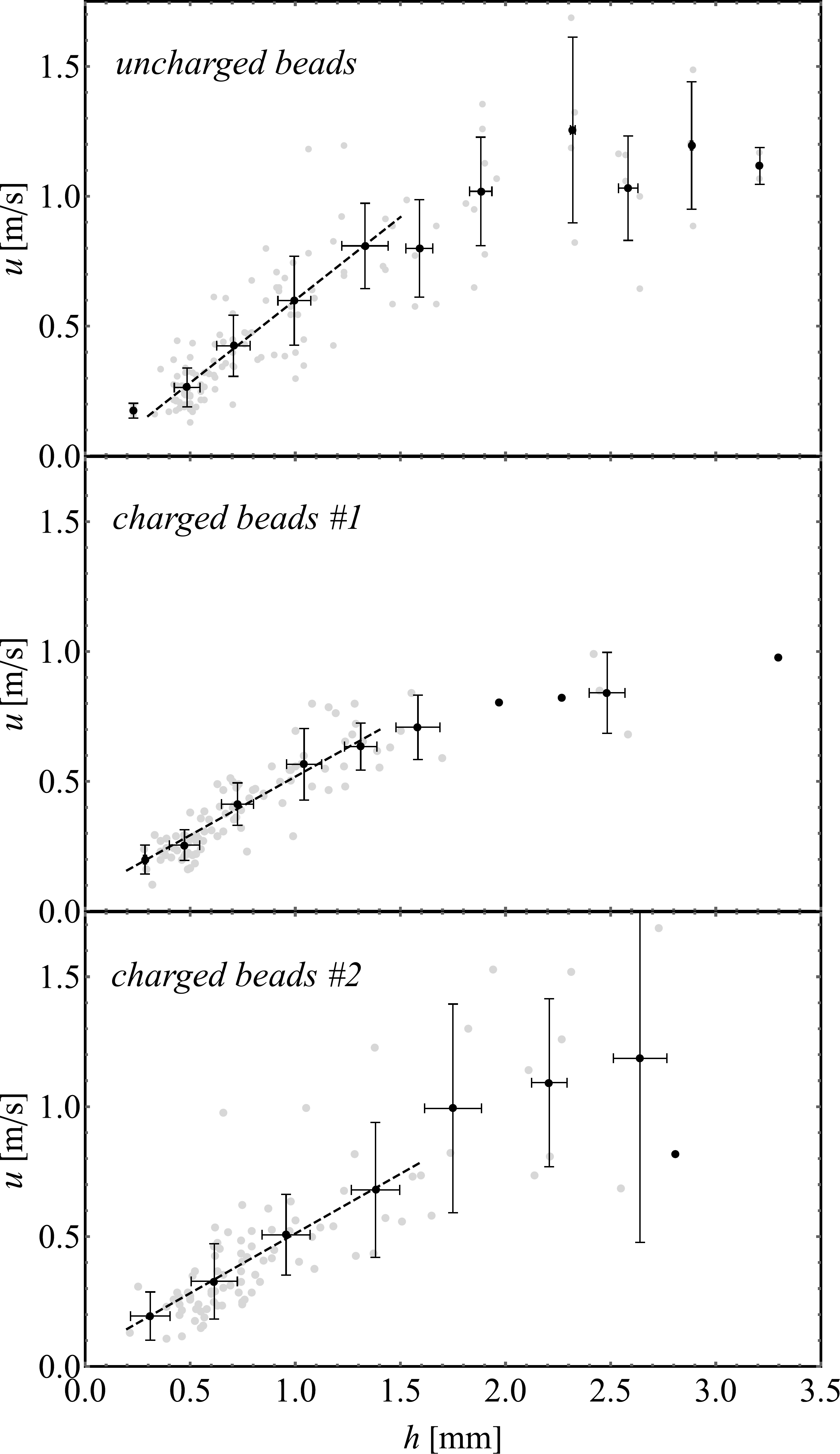}
\caption{Velocity height profile for the three tested conditions in the model system of glass beads. Gray filled circles denote results from applying Eq.~(\ref{eq:x}) to single grain trajectories, while averages and standard deviations are represented by the black circles and error bars, respectively. Dashed lines are fits to the data in the linear regimes.}
\label{fig:daten_wind}
\end{figure}

\begin{figure}
\includegraphics[width=\columnwidth]{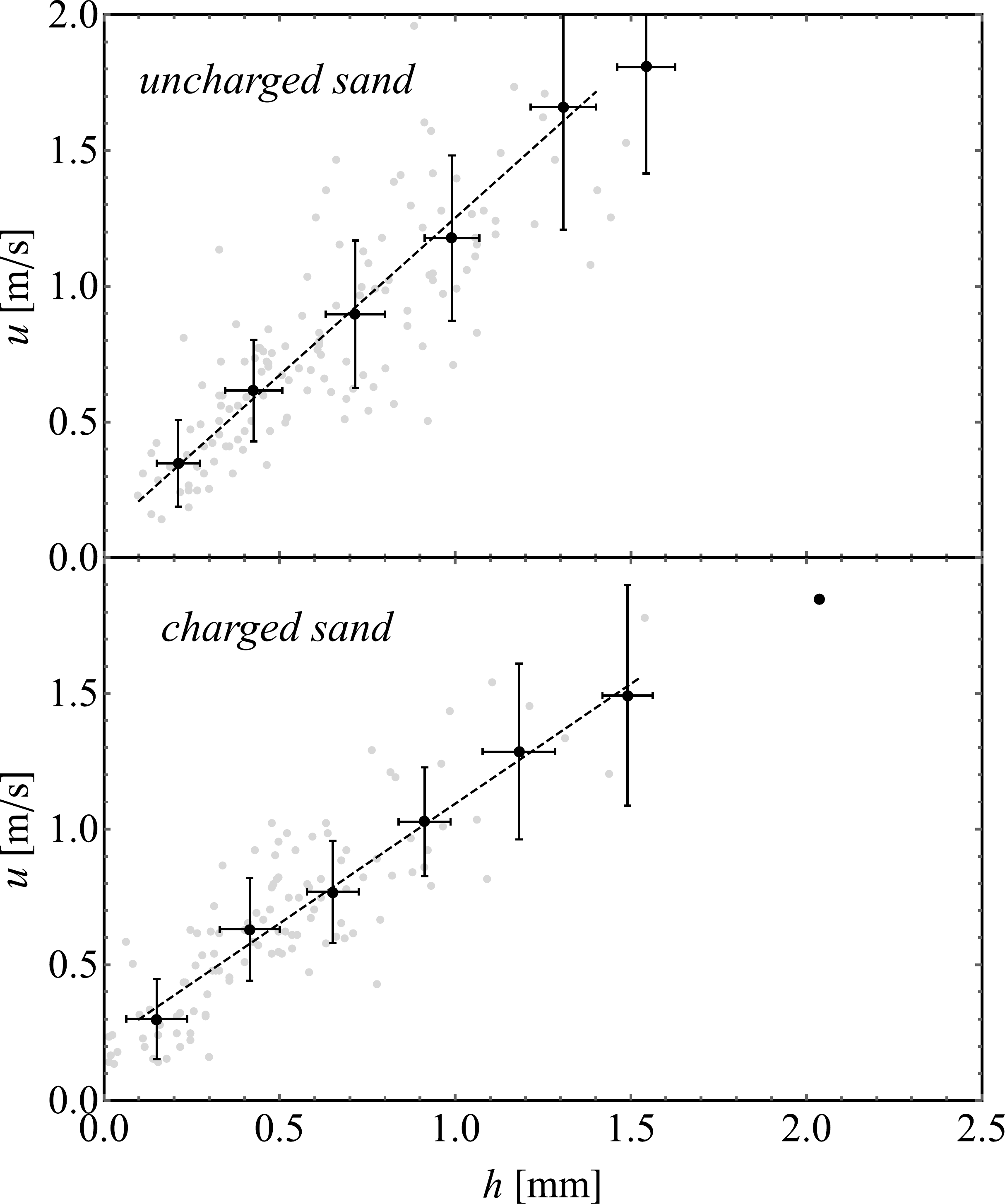}
\caption{Velocity height profile for the uncharged and charged MGS grains. Again, gray filled circles denote results of single grain trajectories, while averages and standard deviations are represented by the black circles and error bars, respectively. Dashed lines are fits to the data in the linear regimes.}
\label{fig:MGSdaten}
\end{figure}

Fig.~\ref{fig:daten_wind} and \ref{fig:MGSdaten} display the wind velocity measured in one experimental run at different heights above the bed for the uncharged and the charged scenarios (see Section \ref{sec:charging}). All experiments have been performed at the minimal threshold flow speed conditions for transport. We see that the mean wind speed increases approximately linearly with the height above the ground (as also found by \cite{Merrison2008}), whereas this linear behavior is valid up to a height of about $1.5\,$mm.

Table~\ref{table} displays the values of the wind velocity gradient, ${{\mathrm{d}}u}/{{\mathrm{d}}h}$, which is the slope of the dashed lines in Fig.~\ref{fig:daten_wind} and \ref{fig:MGSdaten}, and the associated threshold shear velocity  $u_{{\ast}{\mathrm{ft}}}$ obtained from these slopes using Eq.~(\ref{eq:ustar}). As we can see from this table, tribocharging has reduced the value of $u_{{\ast}{\mathrm{ft}}}$ by about 16\,\% in the glass bead scenarios. The realistic analog material shows a similar behavior of a lowered threshold for charged grains. In this case, $u_{{\ast}{\mathrm{ft}}}$ decreases by 12\,\%. The difference in the number of data points results from different measurement times.

In order to ensure reproducibility, all experiments have been conducted twice. Since the threshold value of charged sand is crucial for the application of the results, this measurement has been conducted four times. The maximum deviation from the values reported in Table~\ref{table} was around 0.02\,m/s, which is on the order of the uncertainty of $u_{{\ast}{\mathrm{ft}}}$ resulting from the fits.

\begin{table}
\caption{Wind velocity gradient ${{\mathrm{d}}u}/{{\mathrm{d}}h}$ and wind shear velocity $u_{{\ast}{\mathrm{ft}}}$ at the minimal threshold condition for transport for all experimental scenarios described in Section \ref{sec:charging}.}
\begin{flushleft}
\begin{tabular}{l|c|c|c}
& Data points & $\frac{{\mathrm{d}}u}{{\mathrm{d}}h}$ [s$^{-1}$] & $u_{{\ast}{\mathrm{ft}}}$ [m/s]\\
\hline 
\textit{Uncharged beads} & 190 & $638 \pm 16$ & $0.71 \pm 0.01$ \\
\textit{Charged beads \#1} & 99 & $456 \pm 55$ & $0.60 \pm 0.04$ \\
\textit{Charged beads \#2} & 96 & $459 \pm 16$ & $0.60 \pm 0.01$ \\
\hline
\textit{Uncharged sand} & 140 & $1159 \pm 60$ & $0.91 \pm 0.02$ \\
\textit{Charged sand} & 115 & $882 \pm 38$ & $0.80 \pm 0.02$
\label{table}
\end{tabular}
\end{flushleft}
\end{table}

\section{Model for the minimal threshold wind shear velocity for aerodynamic entrainment}

\begin{figure}
 \centering 
\includegraphics[width=0.6\columnwidth]{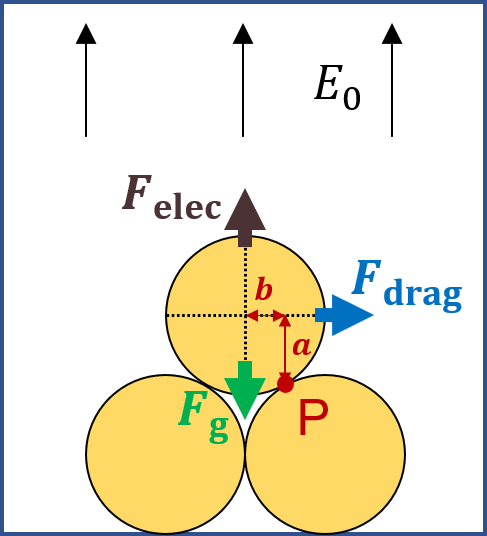}
\caption{Schematical diagram of the main relevant forces acting on a particle that is protruding from the granular bed considered in our experiments. $F_{\mathrm{g}}$, $F_{\mathrm{drag}}$ and $F_{\mathrm{elec}}$ denote the gravitational, drag, and electric forces, respectively, while the distances $a$ and $b$ denote the moment arms of these forces relative to the pivoting point P. Furthermore, $E_0$ denotes the near-surface electric field generated by the granular bed, by considering that the particles have a net positive charge $q_{\mathrm{eff}} > 0$ (net negative charge, $q_{\mathrm{eff}} < 0$, would produce a downward-pointing $E_0$, but $F_{\mathrm{elec}}$ would be still upward-pointing).}
\label{fig:model}
\end{figure}

The goal of this section is to shed further light on the experimentally observed effect of electric charges on the minimum wind shear velocity $u_{{\ast}{\mathrm{ft}}}$ required to initiate transport, by means of modeling. Following \cite{Bagnold1941}, $u_{{\ast}{\mathrm{ft}}}$ can be estimated from the balance of the lifting and resisting torques acting on a particle protruding from the granular bed. Specifically, such a particle will be entrained by the wind flow when it pivots around the point P in Fig.~\ref{fig:model}, which is the point of contact with its supporting neighbor downwind. This balance is computed next both for the uncharged bed and by considering tribocharged particles. The models are based on the measurements with glass beads since these represent well-defined spherical particles.

\subsection{Modeling $u_{{\ast}{\mathrm{ft}}}$ for the uncharged granular bed}

\citet{Bagnold1941} obtained a simple expression for $u_{{\ast}{\mathrm{ft}}}$ by considering the balance between the drag and gravitational torques,
\begin{equation}
aF_{\mathrm{drag}} = bF_{\mathrm{g}} \label{eq:balance}
\end{equation}
where the distances $a$ and $b$, defined in Fig.~\ref{fig:model}, are proportional to the particle diameter $d$, while $F_{\mathrm{drag}}$ and $F_{\mathrm{g}}$ are the drag and gravitational forces, respectively. Furthermore, by neglecting buoyancy, $F_{\mathrm{g}}$ is given by the equation
\begin{equation}
F_{\mathrm{g}} = \frac{\pi}{6}d^3{\rho}_{\mathrm{p}}g \label{eq:Fg}
\end{equation}
where ${\rho}_{\mathrm{p}}$ denotes the particle density and $g=9.81$\,m/s$^{2}$ is gravity, while $F_{\mathrm{drag}}$ can be estimated using the equation
\begin{equation}
F_{\mathrm{drag}} = K_{\mathrm{d}}{\rho}_{\mathrm{f}}d^2u_{\ast}^2 \label{eq:Fdrag}
\end{equation}
with $u_{\ast}$ standing for the wind shear velocity and ${\rho}_{\mathrm{f}}$ denoting the air density. The values of the main parameters can be found in Table~\ref{tab:parameters}. In Eq.~(\ref{eq:Fdrag}), $K_{\mathrm{d}}$ is a dimensionless coefficient that depends on the Reynolds friction number. By combining Eqs.~(\ref{eq:balance}), (\ref{eq:Fg}) and (\ref{eq:Fdrag}), the following expression is obtained for the minimal threshold wind shear velocity:
\begin{equation}
u_{{\ast}{\mathrm{ft}}} = A_{\mathrm{d}}\sqrt{\frac{\rho_{\mathrm{p}}gd}{\rho_{\mathrm{f}}}}  \label{eq:uft}
\end{equation}
where the coefficient $A_{\mathrm{d}}$, called dimensionless threshold friction velocity, encodes the dependence on ${\mbox{Re}}_{{\ast}{\mathrm{ft}}}$, the Reynolds number at the threshold friction velocity.

We note that, in Eq.~(\ref{eq:balance}), we have not included attractive particle interaction forces caused by interatomic or intermolecular (van der Waals) interactions, since it has been shown that the effect of such forces on $u_{{\ast}{\mathrm{ft}}}$ is small for particle sizes larger than 200\,\textmu m on Earth \citep{Demirci2020a,Shao2000,Lu_et_al_2005,Fu_2020}. However, various factors, such as the intermittent behavior of the wind velocity and the thickness of the boundary layer, affect the dependence of $A_{\mathrm{d}}$ on ${\mbox{Re}}_{{\ast}{\mathrm{ft}}}$. The derivation of an expression for $A_{\mathrm{d}}$ that reliably accounts for these factors is still a matter of research \citep{Lu_et_al_2005,Paehtz_et_al_2020,Fu_2020,Swann2020}.

\begin{table}
\caption{\label{tab:parameters}Values of the main parameters of the model.}
\begin{tabular}{l|c|c}
Parameter & Symbol & Value \\
\hline
\hline 
Particle diameter & $d$ & 434\,\textmu m \\
\hline
Particle density & $\rho_{\mathrm{p}}$ & $2600\,$kg/m$^3$ \\
\hline
Air density & ${\rho}_{\mathrm{f}}$ & $0.023\,$kg/m$^3$ \\
\hline
Height of the granular bed & $H_{\mathrm{bed}}$ & $1\,$cm \\
\hline
Bed packing fraction & $\phi$ & $0.60$ \\
\end{tabular}
\end{table}

Therefore, using Eq.~(\ref{eq:uft}), we estimate $A_{\mathrm{d}}$ from the value $u_{{\ast}{\mathrm{ft}}} \approx 0.71\,$m/s obtained in the experiment with the uncharged glass beads, which yields $A_{\mathrm{d}} \approx 0.032$. This value of $A_{\mathrm{d}}$ is, thus, below the value of $A_{\mathrm{d}}$ characteristic of aeolian sand particles in the Earth's atmosphere (about $0.11$). Our finding aligns with recent observations \citep{Swann2020,Fu_2020} of lower-than-previously-thought $u_{{\ast}{\mathrm{ft}}}$ on Mars --- since early models of Martian aeolian processes assumed Earth's value of $A_{\mathrm{d}}\approx0.11$ (for reviews, see \cite{Greeley1985,Kok2012}).

In the next section, we extend our model to elucidate how an electric force decreases the minimal threshold for aeolian sand transport initiation in the experiments.

\subsection{Modeling $u_{{\ast}{\mathrm{ft,elec}}}$ for the charged granular bed}

The physics of electrostatic interactions in wind-blown particle systems is still poorly known \citep{Kok_and_Renno_2006,Zhang_et_al_2015}. Furthermore, the modeling of tribocharging in granular systems, as well as of the electric forces between tribocharged particles, constitutes active matter of research \citep{Lacks2019}. Taking an additional electric force $F_{\rm{elec}}$ into account, the balance of the torques acting on the particle protruding from the granular surface in Fig.~\ref{fig:model} reads
\begin{equation}
aF_{\mathrm{drag}} + bF_{\mathrm{elec}}= bF_{\mathrm{g}} \label{eq:balance_elec}
\end{equation}
where $F_{\mathrm{g}}$ and $F_{\mathrm{drag}}$ are given by Eqs.~(\ref{eq:Fg}) and (\ref{eq:Fdrag}). By substituting these expressions into Eq.~(\ref{eq:balance_elec}), the following equation is obtained for the minimal threshold wind shear velocity for entrainment of the electrified grains,
\begin{equation}
    u_{{\ast}{\mathrm{ft,elec}}} = u_{{\ast}{\mathrm{ft}}}\sqrt{\mbox{max}{\left\{0, \left[1 -\frac{F_{\rm{elec}}}{F_{\rm{g}}}\right]\right\}}}, \label{eq:uft_elec}
\end{equation}
where $u_{{\ast}{\mathrm{ft}}}$, the corresponding threshold shear velocity for the uncharged bed, is given by Eq.~(\ref{eq:uft}). In the following, we present three different approaches to model the electric force on a soil particle and compare the results with the experimental outcome.

\subsubsection{\label{sec:model1}Electric field of random net charges}

The charge measurements revealed that the beads in the granular bed carry net charges that create an electric field at the bed surface. Depending on the sign of the charges and the resulting field, a particle protruding from the surface may feel an additional attracting or repelling force. As the exact charge configuration in the bed is unknown, we take a random charge distribution as a first estimate to evaluate the force induced by net charges.

\begin{figure}
\centering 
\includegraphics[width=1.0\columnwidth]{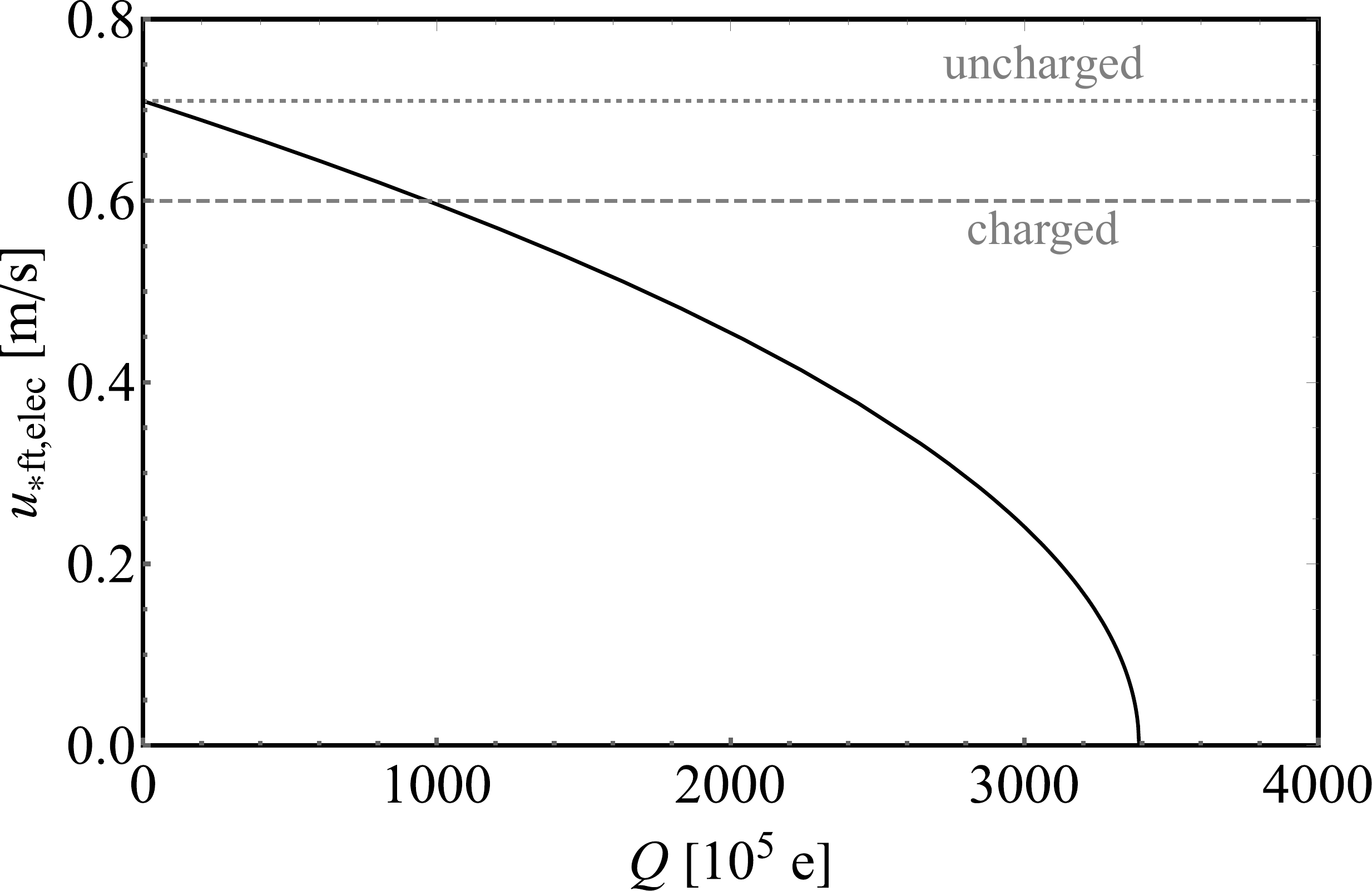}
\caption{Modeled $u_{{\ast}{\mathrm{ft,elec}}}$ as a function of the charge $Q$ of a protruding particle in the electric field of randomly distributed net charges. The horizontal lines indicate the values of the threshold wind velocities obtained in the experiments.}
\label{fig:results_random}
\end{figure}

For simplicity, we model the granular bed based on a simple cubic geometry with the respective dimensions of the experiment. The individual charges $Q_i$ of each bead are assumed to be located in the center of the particle and to be randomly distributed between $-10^6$\,e and $+10^6$\,e, which is an overestimate, considering the measurement in Fig.~\ref{fig:daten_charge}. 
The electric field $\vec{E}_0$ at a surface spot $\vec{r}_{\rm{s}}$ can then be calculated by summing up the contributions of all charges $Q_i$ located at positions $\vec{r}_i$ in the bed:
\begin{equation}
    \vec{E}_0= \frac{1}{4\pi \epsilon_0} \sum_i Q_i\frac{\vec{r}_{\rm{s}} - \vec{r}_i}{{|\vec{r}_{\rm{s}} - \vec{r}_i}|^3}.
\end{equation}
As only the vertical component of the field $E_{\rm{0,z}}$ is relevant to determine the threshold wind velocity, the calculation can be limited to this direction. The maximum upward force $F_{\rm{elec}}$ acting on a surface particle with a charge $Q$ can be calculated according to
\begin{equation}
    \label{eq:Felec_max}
    F_{\rm{elec}}=Q \cdot \mbox{max}(E_{\rm{0,z}}),
\end{equation}
where $\mbox{max}(E_{\rm{0,z}})$ denotes the maximum vertical component of the electric field generated in any surface point. Since the real charge distribution within the bed, as well as on the bed surface, is not known, we treat $Q$ as a model parameter that will be estimated from comparison with the experimental data. Combining Eqs.~(\ref{eq:uft_elec}) and (\ref{eq:Felec_max}) leads to the minimum threshold wind velocity in the field of random charges. Fig.~\ref{fig:results_random} depicts $u_{{\ast}{\mathrm{ft,elec}}}$ in dependence of $Q$ for one simulation run. Moreover, the horizontal dotted and dashed lines denote the experimental mean values of the minimal threshold wind shear velocity obtained for the uncharged bed ($\approx 0.71\,$m/s) and for the charged beds ($\approx 0.60\,$m/s), respectively.

As we can see, the model predicts reduced values for the threshold wind velocity for charged beds. To be consistent with the experimental results, a charge of $Q=10^8$\,e is required, which exceeds the observed charges by several orders of magnitude. We are aware that this model is very simplified regarding the exact geometry and charge distribution of the granular bed. However, from the magnitude of the estimate, we can conclude that randomly distributed net charges are not sufficient to explain the lower thresholds observed in the experiments.

\subsubsection{Electric field above a conducting plane}

If each glass bead is assumed to carry the same effective charge, the influence on the threshold wind velocity may be higher. In the extreme case, we can regard the bed surface as a flat conducting plane. Such a plane generates an electric field $E_0$ of magnitude \citep{Griffiths_1999,Jackson_1999}
\begin{equation}
E_0 = \frac{\sigma}{2{\epsilon}_0},  \label{eq:E0}
\end{equation}
where $\sigma$ is the charge per unit area. Furthermore, we model $\sigma$ using the equation
\begin{equation}
\sigma = \frac{n_{\mathrm{particles}} \cdot q_{\mathrm{eff}}}{A_{\mathrm{bed}}} \label{eq:sigma}
\end{equation}
where $n_{\mathrm{particles}}$ is the total number of particles in the granular bed, $A_{\mathrm{bed}}$ is the bed surface area, and $q_{\mathrm{eff}}$ is the effective (mean) charge of the particles within the bed. Furthermore, $n_{\mathrm{particles}}$ can be estimated with the equation
\begin{equation}
n_{\mathrm{particles}} = \phi \cdot \frac{A_{\mathrm{bed}}H_{\mathrm{bed}}}{(\pi/6)d^3}, \label{eq:nparticles}
\end{equation}
where $H_{\mathrm{bed}}$ is the thickness of the granular bed and $\phi$ its packing fraction. Therefore, $H_{\mathrm{bed}}\,A_{\mathrm{bed}}$ gives the volume of the bed, while the denominator of Eq.~(\ref{eq:nparticles}) denotes the volume of one single particle.

\begin{figure}
\centering 
\includegraphics[width=1.0\columnwidth]{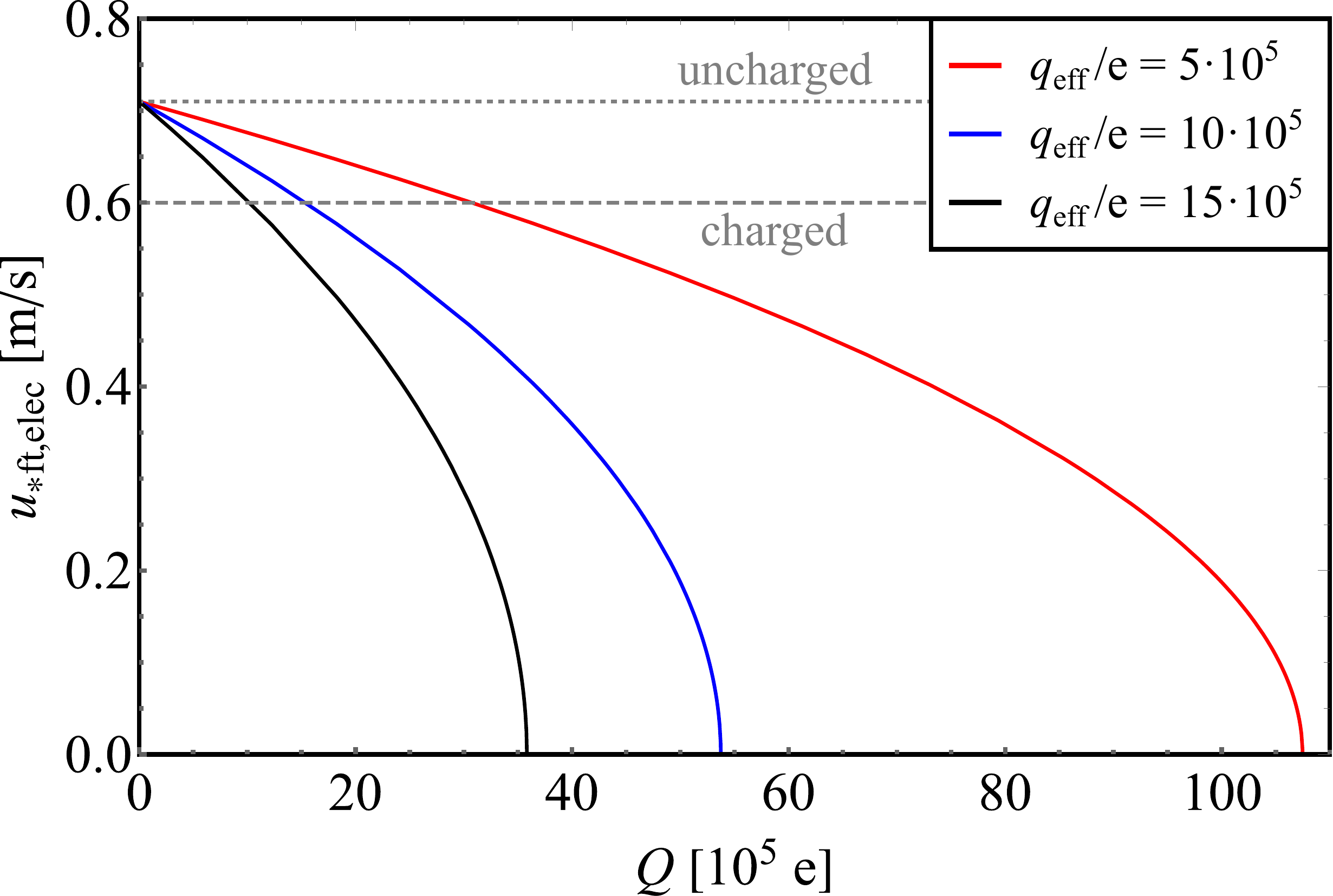}
\caption{Modeled $u_{{\ast}{\mathrm{ft,elec}}}$ as a function of the charge $Q$ of a protruding particle in the electric field of the granular bed for different effective charge magnitudes of the bed particles. The horizontal lines indicate the values of the threshold wind velocities obtained in the experiments.}
\label{fig:results_conducting}
\end{figure}

Just as in the previous section, we assume that the electric force $F_{\mathrm{elec}}$ on the protruding particle in Fig.~\ref{fig:model} can be estimated using the following equation:
\begin{equation}
F_{\mathrm{elec}} = Q \cdot E_0, \label{eq:Felec}
\end{equation}
where the model parameter $Q$ again denotes the characteristic charge value of the protruding surface particle. Figure \ref{fig:results_conducting} shows the predicted values of $u_{{\ast}{\mathrm{ft,elec}}}$ (solid lines), which are calculated using Eqs.~(\ref{eq:uft_elec}) and (\ref{eq:Felec}), as a function of $Q$  for different values of $q_{\mathrm{eff}}$. 

As we can see from Fig.~\ref{fig:results_conducting}, the model would be consistent with the values observed in the experiments, by applying an effective charge magnitude $q_{\mathrm{eff}}$ and charge $Q$ of the surface particle of the order of $15\cdot10^5$\,e. This result is closer to the measured charges in Fig.~\ref{fig:daten_charge} than the first approach in Section~\ref{sec:model1}, but the required charges are still significantly higher than the observed values. Moreover, we note that the effective charge $q_{\mathrm{eff}}$ of the bed particles is overestimated, as the model does not account for the fact that both positive and negative charges occur in the transport layer. For lower $q_{\mathrm{eff}}$, even higher charges $Q$ would be required to explain the experimental observations.

 \subsubsection{Q-patch model}

Electrostatically induced attraction or repulsion between two particles is not only determined by net charges; multipoles also have to be considered \citep{Grosjean2020}. \citet{Jungmann2018} found that the sticking behavior of glass beads can be explained by a shift of the net charge from the center to a near-surface spot. \citet{Steinpilz2020b} considered a complex charge pattern on the surface with patches of negative and positive charge. These Q-patches only measure a few microns but carry a significant amount of charge comparable to the net charge of the particle. 

\begin{figure}
\centering 
\includegraphics[width=1.0\columnwidth]{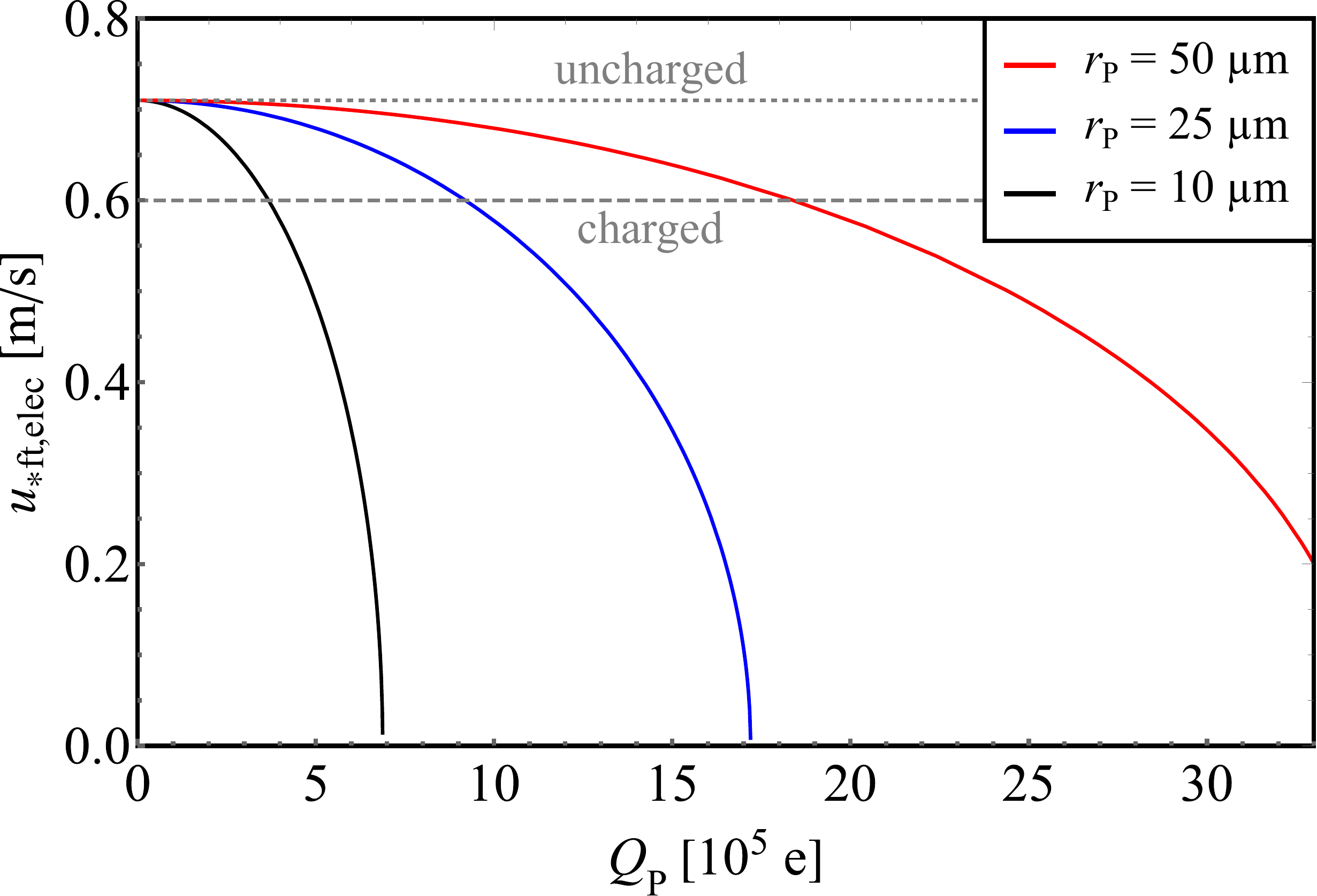}
\caption{Calculated $u_{{\ast}{\mathrm{ft,elec}}}$ as a function of the charge $Q_{\rm{P}}$ of two patches at different distances $r_{\rm{P}}$ applying a Q-patch model. The horizontal lines indicate the values of the threshold wind velocities obtained in the experiments.}
\label{fig:results_q}
\end{figure}

Here we adopt this model and calculate the electric force acting on a protruding particle based on a displaced center of charge. In this case, Coulomb forces can be much stronger, as they decrease with the distance $r$ of two charges according to $1/r^2$. If we assume that two beads in contact interact over two equally charged patches in distance $r_{\rm{P}}$, the Coulomb force reads
\begin{equation}
    \label{eq:Felec_q}
    F_{\rm{elec}}=\frac{1}{4\pi\epsilon_0} \frac{{Q_{\rm{P}}}^2}{{r_{\rm{P}}}^2},
\end{equation}
where $Q_{\rm{P}}$ quantifies the charge of the two patches. By substituting this expression into Eq.~(\ref{eq:uft_elec}), the threshold wind velocities in the Q-patch model can be calculated. Since this model is based on charged patches near the contact point of two beads, the exact geometry of the granular bed is of minor importance to evaluate the lifting forces. Fig.~\ref{fig:results_q} shows the results in dependence of the charge $Q_{\rm{P}}$. As we can see, the charges required to match the experiment are in the range of the measured charges in Fig.~\ref{fig:daten_charge} for a charge separation of 10\,\textmu m.

\section{Discussion and application to Mars}

Our results revealed values much lower than previously thought of the minimal threshold wind shear velocity $u_{{\ast}{\mathrm{ft}}}$ in a Martian environment using the model particles. Specifically, the experiments with an uncharged bed of glass beads yielded $u_{{\ast}{\mathrm{ft}}} \approx 0.71\,$m/s, which is more than 3 times lower than the value of $u_{{\ast}{\mathrm{ft}}}$ predicted by applying Bagnold's equation to the air pressure conditions and particle size and density of our experiments ($\approx 2.41\,$m/s). We have shown that the lower experimental threshold can be explained by a smaller value of the Bagnold coefficient $A_{\mathrm{d}}$ (see Eq.~(\ref{eq:uft})). The threshold friction velocity obtained for the uncharged sample of MGS is in the range of similar analog materials \citep{Musiolik2018, Kruss2020}.

We note that the value of $u_{{\ast}{\mathrm{ft}}}$ assessed in our experiments is the so-called threshold for sporadic, intermittent grain movement, which corresponds to the transition from a static bed to the motion of individual grains being lifted from the bed and performing a few hops over it before coming to a halt. These bursts of grain motion are the precursors of the state of continuous motion, in which saltating grains can be observed continuously over the bed. The threshold for this state of continuous motion has been called the general threshold and has been found to be 10\,\% larger than the threshold for sporadic motion \citep{Swann2020}. Nevertheless, because the bursts of grain hop provide the mechanism underlying the cascading transport that leads to the continuous mode of transport, we follow previous work \citep{Swann2020} and adopt the onset of sporadic bursts as the criterion to define the Martian fluid threshold, $u_{{\ast}{\mathrm{ft}}}$.

Furthermore, we have found that tribocharging can reduce the values of the minimal wind shear velocity required for sediment transport substantially. Not only is this effect observed with a model sample of monodisperse glass beads, but also the MGS measurements show evidence of a lower threshold in a realistic Martian environment. That is, sand grains can be lifted more easily if electric charges are present. We did not investigate erosion rates or the time needed for a potential cascading motion to set in. Since there are also particle configurations where the additional electric force is attractive, the erosion rate at the onset of motion is expected to be lower compared to an uncharged bed at the corresponding threshold friction velocity. Nevertheless, it is beneficial for erosion to already set in at lower wind velocities where the uncharged bed is still at rest just as some wind gusts might initiate saltation at lower wind speeds \citep{Swann2020}.

Based on the glass sample, we developed a model to predict $u_{{\ast}{\mathrm{ft,elec}}}$ under consideration of an electric force (Eq.~(\ref{eq:uft_elec})), for which we tested different approaches. It was shown that, as long as charges are considered to be located in the center of the beads, neither the Coulomb force of randomly distributed net charges nor the electric field of a conducting plane can reproduce the observed values. Based on results of \citet{Jungmann2018}, \citet{Steinpilz2020b} and \citet{Grosjean2020} among others, we also applied a model of charged surface patches. We note that there is no detailed analysis of the surface charge pattern, i.e., the size of the patches and the charges incorporated in them. However, the basic model is the only one that can reproduce the order of magnitude of measured charges. Therefore, we see the repulsion of charged surface patches as the most likely mechanism to explain the observed lifting of glass beads. In fact, similar ideas of charged patches on grains in close contact, though generated by different mechanisms, are currently favored for ejecting dust grains from atmosphereless bodies \citep{Wang2016}.

Our experiments may provide a basis to estimate electric charges on Martian sands, e.g., in terrains where electrostatics may be a problem for landers on Mars \citep{Farrell_et_al_2021}. Furthermore, these insights will be useful to improve models of sand transport on the Martian surface and to predict wind conditions necessary for aeolian geomorphodynamics on Mars \citep{Parteli_and_Herrmann_2007, Paehtz_et_al_2014, Laemmel_et_al_2018}.

Given the large size of the glass beads considered in our study ($d=434$\,\textmu m), van der Waals forces are negligibly small compared to gravitational forces. However, in order to adapt our model to particles such as the used sand, van der Waals forces should be included, since the cohesive interactions owing to these forces become relevant for particles smaller than 70-100\,\textmu m and at reduced gravity \citep{Shao2000, Castellanos_2005, Kok2012, Schmidt_et_al_2020, Demirci2020a}. Moreover, a theoretical approach to more realistic conditions should also include broader size distributions.

\section{Caveats}

The experiments presented here come with some constraints. The side walls of the plastic container in which the glass sample was vibrated were not covered by glass beads. This might change the charging characteristics, as triboelectrification is significantly influenced by the material of the colliding particles \citep{Lacks2019}. Two dissimilar materials tend to carry charges of different sign and lead to broader charge distributions than the interaction of same materials, which is also seen in the asymmetric distributions in Fig.~\ref{fig:daten_charge}. However, even with a coated container and same material contacts, in the case of sand particles, charge separation leads to a significant reduction in the threshold shear stress. Without knowing the charges in detail, their distribution is assumed to be symmetrical and is not biased toward one sign. Nevertheless, the sand measurements show similar evidence of a lower threshold to the glass beads.

Furthermore, the charging process is influenced by the size distribution of the granular material. Same material tribocharging, e.g., typically leads to positive charges on large grains, leaving the smaller ones with negative charges \citep{Waitukaitis2014}. The monodisperse sample used here proved suitable for first experiments and for developing a theoretical model. The MGS sample, though, shows a broad size distribution and is more representative of the Martian surface. Although it is not possible to measure the exact sizes of moving particles owing to constraints of the optical system, it does not seem that there is a favorable grain size for lifting. Particles around 100\,\textmu m in size, as well as larger ones, are observed to be in motion. This finding also provides some validation for the applied Q-patch model, as the effect can be observed independently of the sign of the charges and their detailed distribution. Regardless of the coating of the container, we observe a lower threshold for charged samples. We note that only one size distribution was employed in the experiments and that we did not quantify the charges. Whether different size distributions of the particles affect the charge-induced erosion process is subject to future work. We also have to note that in a full flow, size-dependent charging and size sorting, e.g. lifting smaller grains higher might result in large electric fields, which will have an impact on particle lifting.

In addition to the sample properties, the pressure regime has to be considered. Mars pressure could not be reached with the setup yet, as the wind is not strong enough to lift particles under 20\,mbar. \citet{Wurm2019} showed that the charge gathering on colliding basalt particles is pressure dependent and has a minimum around a few mbar owing to small-scale discharges. According to their data, the width of the charge distribution drops by a factor of around 2 going from 20\,mbar to Martian pressure. The exact influence on the charge $Q_{\rm{P}}$ expected in the charge patches is unknown, though this might reduce the impact of electric forces on the lifting process on Mars. Glass particles, in general, carry less charges than basalt and do not reach charges high enough for pressure-dependent discharges to play a role \citep{Wurm2019, Steinpilz2020a}. It is not clear to what extent this behavior also applies to the sand sample, which is more basaltic.

\section{Conclusion}

In a wind tunnel experiment with tribocharged grains, we intended to evaluate whether tribocharging during saltation might have an influence on particle lift or not. It clearly does, as the threshold wind speed was reduced by more than 10\,\% for both samples. It should be kept in mind that this was the first setup with a number of constraints. Therefore, this number should not be taken as the final face value. It is unlikely that our parameter combination did hit the largest possible reduction factor in shear stress by chance, and in a natural setting the shear stress might be reduced more strongly.

While strong winds are usually needed to initiate saltation on Mars, this might be much easier after grains are charged triboelectrically. This does not {\it a priori} guarantee that particles are always lifted at lower winds, as charging is required first. The immediate conclusion would be that a wind that subsided for a short time can continue saltation at a reduced threshold afterward. How long the ground remains susceptible to this depends on the discharge timescales. We can only speculate on this but at the dry, low-pressure conditions on Mars, this might be hours, maybe more.

In any case, the experiments clearly show a potentially strong role for saltation on Mars.

\section*{acknowledgements}
This project is supported by DLR Space Administration with funds provided by the Federal Ministry for Economic Affairs and Energy (BMWi) under grant numbers DLR 50 WM 1762 and DLR 50 WM 2142. This project has received funding from the European
Union’s Horizon 2020 research and innovation
program under grant agreement No 101004052. EJRP thanks the German Research Foundation for financial support through the Heisenberg Grant, project 434377576. \\

\bibliography{bib}

\begin{thebibliography}{}
\expandafter\ifx\csname natexlab\endcsname\relax\def\natexlab#1{#1}\fi
\providecommand{\url}[1]{\href{#1}{#1}}
\providecommand{\dodoi}[1]{doi:~\href{http://doi.org/#1}{\nolinkurl{#1}}}
\providecommand{\doeprint}[1]{\href{http://ascl.net/#1}{\nolinkurl{http://ascl.net/#1}}}
\providecommand{\doarXiv}[1]{\href{https://arxiv.org/abs/#1}{\nolinkurl{https://arxiv.org/abs/#1}}}

\bibitem[{{Aplin} {et~al.}(2014){Aplin}, {Houghton}, \& {Nicoll}}]{Aplin2014}
{Aplin}, K.~L., {Houghton}, I. M.~P., \& {Nicoll}, K.~A. 2014, arXiv e-prints,
  arXiv:1404.6905.
\newblock \doarXiv{1404.6905}

\bibitem[{Bagnold(1941)}]{Bagnold1941}
Bagnold, R. 1941, The physics of blown sand and desert dunes" (Methuen and Co.
  Ltd., London)

\bibitem[{{Balme} \& {Hagermann}(2006)}]{Balme2006}
{Balme}, M., \& {Hagermann}, A. 2006, \grl, 33, L19S01,
  \dodoi{10.1029/2006GL026819}

\bibitem[{Bila {et~al.}(2020)Bila, Wurm, Onyeagusi, \& Teiser}]{Bila2020}
Bila, T., Wurm, G., Onyeagusi, F.~C., \& Teiser, J. 2020, Icarus, 339, 113569,
  \dodoi{https://doi.org/10.1016/j.icarus.2019.113569}

\bibitem[{Bourke {et~al.}(2019)Bourke, Balme, Lewis, Lorenz, \&
  Parteli}]{Bourke_et_al_2019}
Bourke, M.~C., Balme, M., Lewis, S., Lorenz, R.~D., \& Parteli, E. 2019,
  Planetary Aeolian Geomorphology (John Wiley \& Sons, Ltd), 261--286,
  \dodoi{https://doi.org/10.1002/9781118945650.ch11}

\bibitem[{{Bridges} {et~al.}(2012){Bridges}, {Bourke}, {Geissler}, {Banks},
  {Colon}, {Diniega}, {Golombek}, {Hansen}, {Mattson}, {McEwen}, {Mellon},
  {Stantzos}, \& {Thomson}}]{Bridges2012}
{Bridges}, N.~T., {Bourke}, M.~C., {Geissler}, P.~E., {et~al.} 2012, Geology,
  40, 31, \dodoi{10.1130/G32373.1}

\bibitem[{{Burr} {et~al.}(2020){Burr}, {Sutton}, {Emery}, {Nield}, {Kok},
  {Smith}, \& {Bridges}}]{Burr2020}
{Burr}, D.~M., {Sutton}, S. L.~F., {Emery}, J.~P., {et~al.} 2020, Aeolian
  Research, 45, 100601, \dodoi{10.1016/j.aeolia.2020.100601}

\bibitem[{Cannon {et~al.}(2019)Cannon, Britt, Smith, Fritsche, \&
  Batcheldor}]{Cannon2019}
Cannon, K.~M., Britt, D.~T., Smith, T.~M., Fritsche, R.~F., \& Batcheldor, D.
  2019, Icarus, 317, 470, \dodoi{https://doi.org/10.1016/j.icarus.2018.08.019}

\bibitem[{Castellanos(2005)}]{Castellanos_2005}
Castellanos, A. 2005, Advances in Physics, 54, 263,
  \dodoi{10.1080/17461390500402657}

\bibitem[{{de Beule} {et~al.}(2014){de Beule}, {Wurm}, {Kelling}, {K{\"u}pper},
  {Jankowski}, \& {Teiser}}]{deBeule2014}
{de Beule}, C., {Wurm}, G., {Kelling}, T., {et~al.} 2014, Nature Physics, 10,
  17, \dodoi{10.1038/nphys2821}

\bibitem[{{Demirci} {et~al.}(2020){Demirci}, {Schneider}, {Steinpilz},
  {Bogdan}, {Teiser}, \& {Wurm}}]{Demirci2020a}
{Demirci}, T., {Schneider}, N., {Steinpilz}, T., {et~al.} 2020, \mnras, 493,
  5456, \dodoi{10.1093/mnras/staa607}

\bibitem[{Esposito {et~al.}(2016)Esposito, Molinaro, Popa, Molfese, Cozzolino,
  Marty, Taj-Eddine, Di~Achille, Franzese, Silvestro, \& Ori}]{Esposito2016}
Esposito, F., Molinaro, R., Popa, C.~I., {et~al.} 2016, Geophysical Research
  Letters, 43, 5501, \dodoi{10.1002/2016GL068463}

\bibitem[{Farrell {et~al.}(2021)Farrell, McLain, Marshall, \&
  Wang}]{Farrell_et_al_2021}
Farrell, W.~M., McLain, J.~L., Marshall, J.~R., \& Wang, A. 2021, The Planetary
  Science Journal, 2, 46, \dodoi{10.3847/psj/abe1c3}

\bibitem[{{Fisher} {et~al.}(2005){Fisher}, {Richardson}, {Newman}, {Szwast},
  {Graf}, {Basu}, {Ewald}, {Toigo}, \& {Wilson}}]{Fisher2005}
{Fisher}, J.~A., {Richardson}, M.~I., {Newman}, C.~E., {et~al.} 2005, Journal
  of Geophysical Research (Planets), 110, E03004, \dodoi{10.1029/2003JE002165}

\bibitem[{{Franzese} {et~al.}(2018){Franzese}, {Esposito}, {Lorenz},
  {Silvestro}, {Popa}, {Molinaro}, {Cozzolino}, {Molfese}, {Marty}, \&
  {Deniskina}}]{Franzese2018}
{Franzese}, G., {Esposito}, F., {Lorenz}, R., {et~al.} 2018, Earth and
  Planetary Science Letters, 493, 71, \dodoi{10.1016/j.epsl.2018.04.023}

\bibitem[{{Fu, L.-T.}(2020)}]{Fu_2020}
{Fu, L.-T.} 2020, Icarus, 114225, \dodoi{10.1016/j.icarus.2020.114225}

\bibitem[{Genc {et~al.}(2019)Genc, M{\"o}lleken, Tarasevitch, Utzat, Nienhaus,
  \& M{\"o}ller}]{Genc2019}
Genc, E., M{\"o}lleken, A., Tarasevitch, D., {et~al.} 2019, Review of
  Scientific Instruments, 90, 075115

\bibitem[{{Greeley} {et~al.}(2003){Greeley}, {Balme}, {Iversen}, {Metzger},
  {Mickelson}, {Phoreman}, \& {White}}]{Greeley2003}
{Greeley}, R., {Balme}, M.~R., {Iversen}, J.~D., {et~al.} 2003, Journal of
  Geophysical Research (Planets), 108, 5041, \dodoi{10.1029/2002JE001987}

\bibitem[{Greeley \& Iversen(1985)}]{Greeley1985}
Greeley, R., \& Iversen, J.~D. 1985, New York: Cambridge University Press,
  \dodoi{10.1017/S0016756800035640}

\bibitem[{Griffiths(1999)}]{Griffiths_1999}
Griffiths, D.~J. 1999, Introduction to Electrodynamics, 3rd ed. (Prentice Hall,
  New Jersey)

\bibitem[{{Grosjean} {et~al.}(2020){Grosjean}, {Wald}, {Sobarzo}, \&
  {Waitukaitis}}]{Grosjean2020}
{Grosjean}, G., {Wald}, S., {Sobarzo}, J.~C., \& {Waitukaitis}, S. 2020,
  Physical Review Materials, 4, 082602,
  \dodoi{10.1103/PhysRevMaterials.4.082602}

\bibitem[{{Heyer} {et~al.}(2020){Heyer}, {Raack}, {Hiesinger}, \&
  {Jaumann}}]{Heyer2020}
{Heyer}, T., {Raack}, J., {Hiesinger}, H., \& {Jaumann}, R. 2020, \icarus, 351,
  113951, \dodoi{10.1016/j.icarus.2020.113951}

\bibitem[{Jackson(1999)}]{Jackson_1999}
Jackson, J.~D. 1999, Classical Electrodynamics, 3rd ed. (John Wiley, Hoboken,
  N. H.)

\bibitem[{{Jungmann} {et~al.}(2018){Jungmann}, {Steinpilz}, {Teiser}, \&
  {Wurm}}]{Jungmann2018}
{Jungmann}, F., {Steinpilz}, T., {Teiser}, J., \& {Wurm}, G. 2018, Journal of
  Physics Communications, 2, 095009, \dodoi{10.1088/2399-6528/aad0d2}

\bibitem[{{Jungmann} {et~al.}(2021){Jungmann}, {Bila}, {Kleinert},
  {M{\"o}lleken}, {M{\"o}ller}, {Schmidt}, {Schneider}, {Teiser}, {Utzat},
  {Volkenborn}, \& {Wurm}}]{Jungmann2021}
{Jungmann}, F., {Bila}, T., {Kleinert}, L., {et~al.} 2021, \icarus, 355,
  114127, \dodoi{10.1016/j.icarus.2020.114127}

\bibitem[{{Koester} {et~al.}(2017){Koester}, {Kelling}, {Teiser}, \&
  {Wurm}}]{Koester2017}
{Koester}, M., {Kelling}, T., {Teiser}, J., \& {Wurm}, G. 2017, \apss, 362,
  171, \dodoi{10.1007/s10509-017-3154-4}

\bibitem[{{Koester} \& {Wurm}(2017)}]{Koester2017a}
{Koester}, M., \& {Wurm}, G. 2017, \planss, 145, 9,
  \dodoi{10.1016/j.pss.2017.07.005}

\bibitem[{Kok {et~al.}(2012)Kok, Parteli, Michaels, \& Karam}]{Kok2012}
Kok, J.~F., Parteli, E. J.~R., Michaels, T.~I., \& Karam, D.~B. 2012, Reports
  on Progress in Physics, 75, 106901, \dodoi{10.1088/0034-4885/75/10/106901}

\bibitem[{Kok \& Renno(2006)}]{Kok_and_Renno_2006}
Kok, J.~F., \& Renno, N.~O. 2006, Geophysical Research Letters, 33,
  \dodoi{https://doi.org/10.1029/2006GL026284}

\bibitem[{{Kruss} {et~al.}(2020){Kruss}, {Musiolik}, {Demirci}, {Wurm}, \&
  {Teiser}}]{Kruss2020}
{Kruss}, M., {Musiolik}, G., {Demirci}, T., {Wurm}, G., \& {Teiser}, J. 2020,
  \icarus, 337, 113438, \dodoi{10.1016/j.icarus.2019.113438}

\bibitem[{{Kuepper} \& {Wurm}(2016)}]{Kuepper2016}
{Kuepper}, M., \& {Wurm}, G. 2016, \icarus, 274, 249,
  \dodoi{10.1016/j.icarus.2016.02.049}

\bibitem[{Lacks \& Shinbrot(2019)}]{Lacks2019}
Lacks, D.~J., \& Shinbrot, T. 2019, Nature Reviews Chemistry, 3, 465,
  \dodoi{https://doi.org/10.1038/s41570-019-0115-1}

\bibitem[{L\"ammel {et~al.}(2018)L\"ammel, Meiwald, Yizhaq, Tsoar, Katra, \&
  Kroy}]{Laemmel_et_al_2018}
L\"ammel, M., Meiwald, A., Yizhaq, H., {et~al.} 2018, Nature Physics, 14, 759,
  \dodoi{10.1038/s41567-018-0106-z}

\bibitem[{{Liuzzi} {et~al.}(2020){Liuzzi}, {Villanueva}, {Crismani}, {Smith},
  {Mumma}, {Daerden}, {Aoki}, {Vand aele}, {Clancy}, {Erwin}, {Thomas},
  {Ristic}, {Lopez-Moreno}, {Bellucci}, \& {Patel}}]{Liuzzi2020}
{Liuzzi}, G., {Villanueva}, G.~L., {Crismani}, M. M.~J., {et~al.} 2020, Journal
  of Geophysical Research (Planets), 125, e06250, \dodoi{10.1029/2019JE006250}

\bibitem[{Lu {et~al.}(2005)Lu, Raupach, \& Richards}]{Lu_et_al_2005}
Lu, H., Raupach, M.~R., \& Richards, K.~S. 2005, Journal of Geophysical
  Research, 110, D24114, \dodoi{10.1029/2005JD006418}

\bibitem[{{M{\'e}ndez Harper} \& {Dufek}(2016)}]{Harper2016}
{M{\'e}ndez Harper}, J., \& {Dufek}, J. 2016, Journal of Geophysical Research
  (Atmospheres), 121, 8209, \dodoi{10.1002/2015JD024275}

\bibitem[{{M{\'e}ndez Harper} {et~al.}(2017){M{\'e}ndez Harper}, {McDonald},
  {Dufek}, {Malaska}, {Burr}, {Hayes}, {McAdams}, \& {Wray}}]{Harper2017}
{M{\'e}ndez Harper}, J.~S., {McDonald}, G.~D., {Dufek}, J., {et~al.} 2017,
  Nature Geoscience, 10, 260, \dodoi{10.1038/ngeo2921}

\bibitem[{Merrison {et~al.}(2008)Merrison, Bechtold, Gunnlaugsson, Jensen,
  Kinch, Nornberg, \& Rasmussen}]{Merrison2008}
Merrison, J., Bechtold, H., Gunnlaugsson, H., {et~al.} 2008, Planetary and
  Space Science, 56, 426 , \dodoi{https://doi.org/10.1016/j.pss.2007.11.007}

\bibitem[{{Merrison} {et~al.}(2007){Merrison}, {Gunnlaugsson}, {N{\o}rnberg},
  {Jensen}, \& {Rasmussen}}]{Merrison2007}
{Merrison}, J.~P., {Gunnlaugsson}, H.~P., {N{\o}rnberg}, P., {Jensen}, A.~E.,
  \& {Rasmussen}, K.~R. 2007, \icarus, 191, 568,
  \dodoi{10.1016/j.icarus.2007.04.035}

\bibitem[{Musiolik {et~al.}(2018)Musiolik, Kruss, Demirci, Schrinski, Teiser,
  Daerden, Smith, Neary, \& Wurm}]{Musiolik2018}
Musiolik, G., Kruss, M., Demirci, T., {et~al.} 2018, Icarus, 306, 25 ,
  \dodoi{https://doi.org/10.1016/j.icarus.2018.01.007}

\bibitem[{{Méndez Harper} {et~al.}(2021){Méndez Harper}, Dufek, \&
  McDonald}]{Harper2021}
{Méndez Harper}, J., Dufek, J., \& McDonald, G.~D. 2021, Icarus, 357, 114268,
  \dodoi{https://doi.org/10.1016/j.icarus.2020.114268}

\bibitem[{{Neakrase} {et~al.}(2016){Neakrase}, {Balme}, {Esposito}, {Kelling},
  {Klose}, {Kok}, {Marticorena}, {Merrison}, {Patel}, \& {Wurm}}]{Neakrase2016}
{Neakrase}, L.~D.~V., {Balme}, M.~R., {Esposito}, F., {et~al.} 2016, \ssr, 203,
  347, \dodoi{10.1007/s11214-016-0296-6}

\bibitem[{P\"ahtz {et~al.}(2020)P\"ahtz, Clark, Valyrakis, \&
  Dur\'an}]{Paehtz_et_al_2020}
P\"ahtz, T., Clark, A.~H., Valyrakis, M., \& Dur\'an, O. 2020, Reviews of
  Geophysics, 58, e2019RG000679, \dodoi{https://doi.org/10.1029/2019RG000679}

\bibitem[{P\"ahtz {et~al.}(2010)P\"ahtz, Herrmann, \&
  Shinbrot}]{Paehtz_et_al_2010}
P\"ahtz, T., Herrmann, H.~J., \& Shinbrot, T. 2010, Nature Physics, 6, 364,
  \dodoi{https://doi.org/10.1038/nphys1631}

\bibitem[{P\"ahtz {et~al.}(2014)P\"ahtz, Parteli, Kok, \&
  Herrmann}]{Paehtz_et_al_2014}
P\"ahtz, T., Parteli, E. J.~R., Kok, J.~F., \& Herrmann, H.~J. 2014, Phys. Rev.
  E, 89, 052213, \dodoi{10.1103/PhysRevE.89.052213}

\bibitem[{Parteli \& Herrmann(2007)}]{Parteli_and_Herrmann_2007}
Parteli, E. J.~R., \& Herrmann, H.~J. 2007, Phys. Rev. Lett., 98, 198001,
  \dodoi{10.1103/PhysRevLett.98.198001}

\bibitem[{Rasmussen {et~al.}(2009)Rasmussen, Kok, \& Merrison}]{Rasmussen2009}
Rasmussen, K.~R., Kok, J.~F., \& Merrison, J.~P. 2009, Planetary and Space
  Science, 57, 804 , \dodoi{https://doi.org/10.1016/j.pss.2009.03.001}

\bibitem[{Rasmussen {et~al.}(2015)Rasmussen, Valance, \&
  Merrison}]{Rasmussen2015}
Rasmussen, K.~R., Valance, A., \& Merrison, J. 2015, Geomorphology, 244, 74 ,
  \dodoi{https://doi.org/10.1016/j.geomorph.2015.03.041}

\bibitem[{{Schmidt} {et~al.}(1998){Schmidt}, {Schmidt}, \&
  {Dent}}]{Schmidt1998}
{Schmidt}, D.~S., {Schmidt}, R.~A., \& {Dent}, J.~D. 1998, \jgr, 103, 8997,
  \dodoi{10.1029/98JD00278}

\bibitem[{{Schmidt} {et~al.}(2017){Schmidt}, {Andrieu}, {Costard}, {Kocifaj},
  \& {Meresescu}}]{Schmidt2017}
{Schmidt}, F., {Andrieu}, F., {Costard}, F., {Kocifaj}, M., \& {Meresescu},
  A.~G. 2017, Nature Geoscience, 10, 270, \dodoi{10.1038/ngeo2917}

\bibitem[{Schmidt {et~al.}(2020)Schmidt, Parteli, Uhlmann, Wörlein, Wirth,
  Pöschel, \& Peukert}]{Schmidt_et_al_2020}
Schmidt, J., Parteli, E.~J., Uhlmann, N., {et~al.} 2020, Advanced Powder
  Technology, 31, 2293, \dodoi{https://doi.org/10.1016/j.apt.2020.03.018}

\bibitem[{Shao \& Lu(2000)}]{Shao2000}
Shao, Y., \& Lu, H. 2000, Journal of Geophysical Research: Atmospheres, 105,
  22437, \dodoi{10.1029/2000JD900304}

\bibitem[{Steinpilz {et~al.}(2020a)Steinpilz, Joeris, Jungmann, Wolf, Brendel,
  Teiser, Shinbrot, \& Wurm}]{Steinpilz2020a}
Steinpilz, T., Joeris, K., Jungmann, F., {et~al.} 2020a, Nature Physics, 16,
  225, \dodoi{https://doi.org/10.1038/s41567-019-0728-9}

\bibitem[{{Steinpilz} {et~al.}(2020b){Steinpilz}, {Jungmann}, {Joeris},
  {Teiser}, \& {Wurm}}]{Steinpilz2020b}
{Steinpilz}, T., {Jungmann}, F., {Joeris}, K., {Teiser}, J., \& {Wurm}, G.
  2020b, New Journal of Physics, 22, 093025, \dodoi{10.1088/1367-2630/abae43}

\bibitem[{{Swann} {et~al.}(2020){Swann}, {Sherman}, \& {Ewing}}]{Swann2020}
{Swann}, C., {Sherman}, D.~J., \& {Ewing}, R.~C. 2020, \grl, 47, e84484,
  \dodoi{10.1029/2019GL084484}

\bibitem[{Teiser {et~al.}(2021)Teiser, Kruss, Jungmann, \& Wurm}]{Teiser2021}
Teiser, J., Kruss, M., Jungmann, F., \& Wurm, G. 2021, The Astrophysical
  Journal, 908, L22, \dodoi{10.3847/2041-8213/abddc2}

\bibitem[{Tinevez {et~al.}(2017)Tinevez, Perry, Schindelin, Hoopes, Reynolds,
  Laplantine, Bednarek, Shorte, \& Eliceiri}]{Tivenez2017}
Tinevez, J.-Y., Perry, N., Schindelin, J., {et~al.} 2017, Methods, 115, 80,
  \dodoi{https://doi.org/10.1016/j.ymeth.2016.09.016}

\bibitem[{{Waitukaitis} {et~al.}(2014){Waitukaitis}, {Lee}, {Pierson},
  {Forman}, \& {Jaeger}}]{Waitukaitis2014}
{Waitukaitis}, S.~R., {Lee}, V., {Pierson}, J.~M., {Forman}, S.~L., \&
  {Jaeger}, H.~M. 2014, Physical Review Letters, 112, 218001,
  \dodoi{10.1103/PhysRevLett.112.218001}

\bibitem[{{Wang} {et~al.}(2016){Wang}, {Schwan}, {Hsu}, {Gr{\"u}n}, \&
  {Hor{\'a}nyi}}]{Wang2016}
{Wang}, X., {Schwan}, J., {Hsu}, H.~W., {Gr{\"u}n}, E., \& {Hor{\'a}nyi}, M.
  2016, \grl, 43, 6103, \dodoi{10.1002/2016GL069491}

\bibitem[{White {et~al.}(1987)White, Greeley, Leach, \& Iversen}]{White1987}
White, B., Greeley, R., Leach, R., \& Iversen, J. 1987, 25th AIAA Aerospace
  Sciences Meeting, 621, \dodoi{10.2514/6.1987-621}

\bibitem[{Wurm {et~al.}(2001)Wurm, Blum, \& Colwell}]{Wurm2001}
Wurm, G., Blum, J., \& Colwell, J.~E. 2001, Icarus, 151, 318 ,
  \dodoi{https://doi.org/10.1006/icar.2001.6620}

\bibitem[{{Wurm} {et~al.}(2019){Wurm}, {Schmidt}, {Steinpilz}, {Boden}, \&
  {Teiser}}]{Wurm2019}
{Wurm}, G., {Schmidt}, L., {Steinpilz}, T., {Boden}, L., \& {Teiser}, J. 2019,
  \icarus, 331, 103, \dodoi{10.1016/j.icarus.2019.05.004}

\bibitem[{{Zhang} \& {Zhou}(2020)}]{Zhang2020}
{Zhang}, H., \& {Zhou}, Y.-H. 2020, Nature Communications, 11, 5072,
  \dodoi{10.1038/s41467-020-18759-0}

\bibitem[{Zhang {et~al.}(2015)Zhang, P\"ahtz, Liu, Wang, Zhang, Shen, Ji, \&
  Cai}]{Zhang_et_al_2015}
Zhang, Y., P\"ahtz, T., Liu, Y., {et~al.} 2015, Phys. Rev. X, 5, 011002,
  \dodoi{10.1103/PhysRevX.5.011002}

\bibitem[{Zheng(2013)}]{Zheng2013}
Zheng, X.-J. 2013, The European Physical Journal E, 36, 138,
  \dodoi{10.1140/epje/i2013-13138-4}

\end{thebibliography}

\end{document}